\documentclass[fleqn,11pt]{article}
\usepackage{graphicx,amssymb}

\newcommand{\AmS}{{\protect\the\textfont2
  A\kern-.1667em\lower.5ex\hbox{M}\kern-.125emS}}
\textheight by \topskip \textwidth 455pt \hfuzz=15pt
  \normalsize
\begin{document}
{\bf LATERAL DISTRIBUTION FOR ALIGNED EVENTS 
  IN MUON GROUPS DEEP UNDERGROUND}\footnote%
{Talk given at the ISVHECRI'2006, Weihai, China}\\
\\
\baselineskip=12pt \leftline{ A.L.Tsyabuk\footnote{e-mail:
atsyabuk@sci.lebedev.ru}, R.A. Mukhamedshin and Yu.V. Stenkin}\\
\\
\baselineskip=12pt \leftline{\footnotesize\it Institute for
Nuclear research of Russian Academy of Sciences}
\baselineskip=10pt \leftline{\footnotesize\it 60th October anniv.
prospect, 7a, Moscow 117312, Russia}
\begin{abstract}
The paper concerns the so-called aligned events observed in cosmic rays. The phenomenon of the
alignment of the most energetic subcores of gamma-ray--hadron ($\gamma-h$) families (particles of
the highest energies in the central EAS core) was firstly found in the "Pamir" emulsion chamber
experiment and related to a coplanar particle production at $E_0>10^{16}$ eV. Here a separation
distribution (distances between pairs of muons) for aligned events has been analyzed throughout
muon groups measured by Baksan Underground Scintillation Telescope (BUST) for threshold energies
$0.85 \div 3.2$ TeV during a period of 7.7 years.  Only muon groups of multiplicity $m\geq 4$ with
inclined trajectories for an interval of zenith angles $50^\circ - 60^\circ$ were selected for the
analysis. The analysis has revealed that the distribution complies with the exponential law.
Meanwhile the distributions become steeper with the increase of threshold energy. There has been
no difference between the lateral distribution of all the groups and the distribution of the
aligned groups.
\end{abstract}


\section{INTRODUCTION}
\label{introduc}

The most essential problem of cosmic ray physics is the study of the energy spectrum of primary
cosmic radiation (PCR), their chemical composition and interactionS of PCR particles with
atmospheric nuclei at superhigh energies, inaccessible for modern accelerators. Muon groups in
underground experiments, being one of the components of extensive air showers (EAS), convey
information about characteristics of cosmic rays. Alongside with these data on muon groups embrace
the interval of primary energies located between direct and indirect methods of measuring: from
$10^{13}$ to $10^{17}$ eV, i.e., they cover the region of the presumed knee of the primary
cosmic-ray spectrum.

Besides, an interesting cosmic-ray phenomenon related to the so-called alignment of clusters of
close particles or relatively isolated particles is also observed in the knee range. This
phenomenon has been first found in experiments on study of gamma-ray--hadron ($\gamma-h$) families
(groups of the most energetic particles in the EAS cores) with the use of X-ray emulsion chambers
(XREC) in events with an observed energy $\Sigma E_\gamma > 700$ TeV \cite{Pamir4_1,Kopenetal}. The
alignment is realized as the tendency of the location of particles near a straight line on the
target diagram that corresponds in the three-dimensional case to the absence of the azimuthal
symmetry with respect to the EAS axis and concentration of particles' momenta near some plane. The
explanation of the whole block of data related to the alignment phenomenon by fluctuations of
cascade development in the atmosphere seems to be improbable $(\ll 10^{-20})$ \cite{JHEP05_2005}.
The influence of the Earth's magnetic field and thunderstorm-cloud electric fields is also
negligible \cite{8ISVHECRIMukh}. It was assumed that the cause of the alignment is trivial
diffraction kinematics effects \cite{Smirnova,Zhu,capdev}; generation of quark-gluon strings
\cite{Halzen}; appearance of a new strong interaction and generation of new super-heavy quarks of
higher color symmetry \cite{White}; semihard double diffraction dissociation with a tension of a
quark-gluon string between a semihardly scattered constituent quark and spectator quarks of the
projectile hadron and its following rupture with creation of secondary particles \cite{Royzen}.

\section{EXPERIMENT}

\begin{figure}[!bth]
\hspace{-0.75pc}
\begin{center}
         \includegraphics[height=14.0pc]{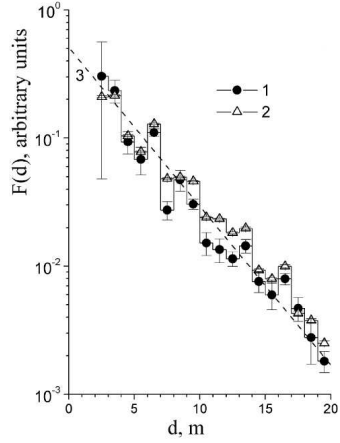}
\caption{Experimental $\,$ lateral $\,$ distributions: 1 -- $\lambda < 0.6$;  2 -- $\lambda > 0.6$;
 3 -- exponential fit ($d_0=3.5$ m) at $E_{thr}^\mu = 0.85$ TeV.}
\end{center}
\label{plot1}
\end{figure}
\begin{figure}[!bth]
\hspace{-0.75pc}
\begin{center}
         \includegraphics[height=14.0pc]{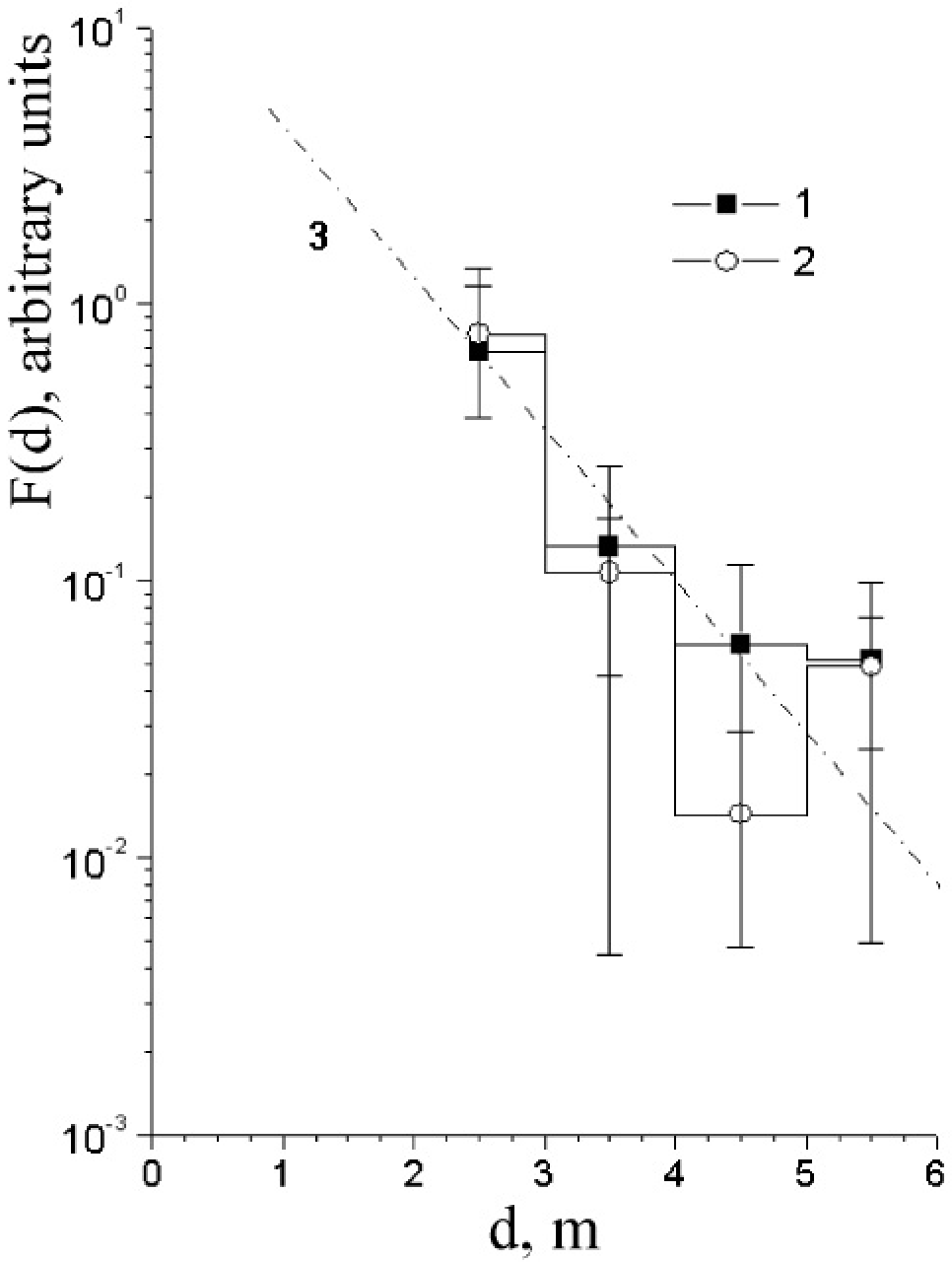}
\caption{Experimental $\,$ lateral $\,$ distributions:
 1 -- $\lambda < 0.6$; 2 -- $\lambda > 0.6$;
 3 -- exponential fit ($d_0=3.5$ m) at $E_{thr}^\mu = 3.2$ TeV.}
\end{center}
\label{plot2}
\end{figure}
\begin{figure}[!tbh]
\hspace{-0.75pc}
\begin{center}
    \includegraphics[height=22pc]{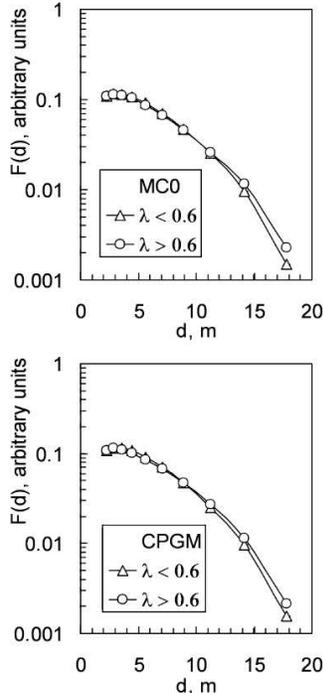}
\vspace{-1.0pc} \caption{Results of Monte Carlo simulations by MC0
and CPGM models at $E_{thr}^\mu = 1$ TeV.}
 \end{center}
\label{fig2}
\end{figure}
There are very few experimental works on aligned events study in
muon groups deep underground: MACRO \cite{sioli} and BUST
\cite{npb06v5p473,29ICRCv9p243}. A search for aligned events has
been done using the alignment parameter \cite{Pamir4_1}:\\
$\lambda_m ={\sum\limits_{i\ne j \ne k}^N cos 2\varphi^k_{ij}}/[
{m(m-1)(m-2)}], $ where $\varphi^k_{ij}$ is the angle between the
lines connecting the $k$-th particle with the $i$-th and $j$-th
particles, $m$ is the  number of particles $(m \geq 4)$.  Note
that  $\lambda_m$ decreases from $\lambda_m=1$ for completely
aligned events to $\lambda_m \simeq -1/[m(m-1)(m-2)]$ in the case
of isotropic events. In \cite{sioli} the distribution of events as
function of parameter $\lambda_m$ is peaked at $\lambda_m =1$ due
to the asymmetry of MACRO detector  geometry $(12\times 76$
m$^2$). Two variants of simulation (CORSIKA/QGSJET and
HEMAS/DPMJET) were made. The experimental data not contradict the
simulation. Results of our analysis of BUST experimental data are
shown in Figs. \ref{plot1} and \ref{plot2} for two muon threshold
energies $E_{thr}^\mu$ (0.85 and 3.2 TeV). One can see that there
is no a significant difference in the lateral distribution for
events selected by $\lambda$ parameter. Note that visible change
of slope in \ref{plot2} for both types of events can be explained
by the admixture of $\sim 10$\% wrong events corresponding to
lower energy threshold due to errors in angular measurements and
very high gradient of the mountain profile.

\section{CONCLUSION}

It was shown in our previous works \cite{npb06v5p473,29ICRCv9p243} that aligned muon events
observed in our experiment are pure statistical ones. In this work we shown that there are no
statistically significant differences in lateral distributions of "aligned" and all events. The
latter confirms once again that our "aligned" muon groups are just a statistical tail of a big
number of events recorded by BUST.

Fig. \ref{fig2} shows the absence of a significant difference in
lateral distributions of muon groups simulated with a QGSJET-like
MC0  model and a coplanar particle generation model (CPGM)
\cite{JHEP05_2005}.  A key for explanation of this result can be
found if one takes into account that a fraction of muons produced
by decaying high-energy mesons generated in the first interaction
is negligible as compared to muons produced in following
generations deeper in the atmosphere.

\vspace{0.5mm} This work is supported by Russian Foundation for Basic Research (projects
04-02-17083, 05-02-17599, 05-02-16781,  06-02-16606, 06-02-16969),  Russian Ministry of Education
and Science (projects SS-5573.2006.2 and SS-4580.2006.02) and Russian Federal Agency of science
and innovations (state contract 02.445.11.7070).


\begin{thebibliography}{9}
\bibitem{Pamir4_1}
Pamir Collaboration,  Proc. 4th ISVHECRI, Beijing (1986) 429.

\bibitem{Kopenetal}
V.V. Kopenkin {\em et al}., Phys.Rev.D 52 (1995) No. 5, 2766.

\bibitem{JHEP05_2005}
R.A. Mukhamedshin, J. High Energy Phys. 05 (2005) 049.

\bibitem{8ISVHECRIMukh}
R.A. Mukhamedshin, Proc. 8th ISVHECRI, Tokyo (1994)  57.

\bibitem{Smirnova}
M.D. Smirnova and Yu.A. Smorodin, Proc. 21st ICRC, Adelaide, 8 (1990) 310.

\bibitem{Zhu}
Q.-Q. Zhu, Y.D. He, and A.X. Huo, Proc. 21st ICRC, Adelaide, 8 (1990) 306.

\bibitem{capdev}
J.N. Capdevielle  {\em et al}., Proc. 27th ICRC, Hamburg, 1 (2001)  1410.

\bibitem{Halzen}
F. Halzen and D.A. Morris, Phys.Rev.D 42 (1990) 1435.

\bibitem{White}
A.R. White, Int.J.Mod.Phys.A8 (1993) 4755.

\bibitem{Royzen}
I.I. Royzen, Mod. Phys. Lett. A9 (1994) No. 38 3517.

\bibitem{sioli}
M. Sioli, PhD thesis. Univ. Degli Studi di Bologna. 1999.

\bibitem{npb06v5p473}
A.L. Tsyabuk {\em et al}., Nucl. Phys. B (Proc Suppl.) 151 (2006) 473.

\bibitem{29ICRCv9p243}
A.L. Tsyabuk {\em et al}., Proc. 29th ICRC Pune, 9 (2005) 243

\end{thebibliography}
\end{document}